\documentclass[11pt]{iopart}
\usepackage{iopams,graphicx,url}
\expandafter\let\csname equation*\endcsname\relax
\expandafter\let\csname endequation*\endcsname\relax

\begin{document}

\title{Spacetime, Spin and Gravity Probe B}

\author{J.M. Overduin,$^{1,2}$}
\address{$^1$ Department of Physics, Astronomy and Geosciences, Towson University, Towson, MD 21252}
\address{$^2$ Department of Physics and Astronomy, Johns Hopkins University, Baltimore, MD 21218}
\ead{joverduin@towson.edu}

\begin{abstract}
It is more important than ever to push experimental tests of gravitational theory to the limits of existing technology in both range and sensitivity. 
This brief review focuses on spin-based tests of General Relativity and their implications for alternative, mostly non-metric theories of gravity motivated by the challenge of unification with the Standard Model of particle physics.
The successful detection of geodetic precession and frame-dragging by Gravity Probe B places new constraints on a number of these theories, and increases our confidence in the theoretical mechanisms underpinning current ideas in astrophysics and cosmology.
\end{abstract}

\maketitle

\section{Introduction}

The need to expand the scope and sensitivity of gravitational tests has only grown more pressing in the years since Gravity Probe B was conceived.
Our best theory of gravity, General Relativity (GR), has survived a century of experimental tests, but with the exception of the binary pulsar, these have been limited to weak fields and low velocities.
GR cannot explain observations on cosmological scales unless the matter in the Universe is supplemented with vast and finely-tuned quantities of unseen dark matter and energy.
The theory predicts its own demise in the form of spacetime singularities inside collapsed stars and at the cores of most galaxies.
Related to this is perhaps the most pressing problem of all, the challenge of reconciling the geometrical picture of gravitation inherent in Einstein's theory with the other interactions of physics, as successfully described by the Standard Model of particle physics (SM).
In the past, unification of disparate branches of physics has repeatedly cleared the way for tremendous progress, not only in scientific understanding but eventually in society as well.
%This progress has largely stalled since the successes of electroweak unification in the 1960s, as recently validated by the experimental confirmation of the Higgs particle.
For fifty years there has been much theoretical activity, but little widely agreed-upon progress in bridging the largest remaining fissure in the foundation of physics: that separating gravity from everything else.
The needed breakthrough is likely to come from experiment.

\section{Gravity}

Experiments in gravitation can usefully be divided into those that test the theories (including GR) and those that test the underlying principles.
Underlying principles include such basic axioms as local position invariance (LPI) and local Lorentz invariance (LLI).
The principle of most direct physical relevance to GR is the equivalence principle (EP).
An increase in sensitivity of five orders of magnitude over existing EP experiments, if coupled with a sufficiently diverse set of test-mass materials, could have unprecedented potential to either point the way toward a successful incorporation of gravity into the SM, or to establish % that the gravitational couplings of any new fields of the kind generically predicted by unified theories must be unnaturally weak \cite{STEP}.
that any such unification necessarily involves an unnatural degree of fine-tuning in coupling strengths \cite{STEP}.

Tests of GR were inaugurated by the three ``classical tests,'' gravitational redshift, light deflection, and the perihelion shift of Mercury.
The gravitational redshift effect, which slows or ``warps'' time in a gravitational field, follows from the EP alone so is sometimes classified with tests of underlying principles.
It was confirmed to high precision by Gravity Probe~A in 1976, among other experiments, and is nowadays reconfirmed with every routine use of a Global Positioning Satellite (GPS) receiver.
Light deflection in the field of the Sun has been found using radio interferometry to agree with GR at the 0.01\% level, and Sloan Digital Sky Survey data on gravitational lensing by distant galaxies have confirmed the same effect there at the 10\% level.
Mercury's perihelion shift agrees with GR to within the 5\% uncertainty currently associated with the solar quadrupole moment, and the analogous effect for the LAGEOS~II satellite around the Earth has been confirmed at 2\% \cite{Will}.

The fourth and fifth tests of GR involved radar time delay (or the Shapiro effect) and loss of energy due to the emission of gravitational waves from binary pulsar systems.
Time delay of signals to and from the Cassini spacecraft currently provides GR with its strongest empirical support, agreeing with the theory at the 0.001\% level.
One-way time delay has also been measured in the light from a millisecond pulsar.
The binary pulsar discovered by Hulse and Taylor in 1974 loses energy at a rate that is consistent with GR at the 0.02\% level.
Other binary systems have now yielded valuable constraints as well, including a double pulsar whose consistency with GR can be checked in five independent ways (including geodetic precession).
Direct detection of gravitational waves from such systems may soon provide an entirely new test of Einstein's theory; null results from LIGO and Virgo are already encroaching on the astrophysical parameter space for binary black-hole mergers within 300~Mpc \cite{Gwaves}.

The primary objective of the Gravity Probe~B (GP-B) mission was to conduct a sixth and seventh test of GR through detection of the geodetic effect and the frame-dragging (or Lense-Thirring) effect.
%(As auxiliary benefits, the experiment was also planned to yield new measurements of gravitational redshift, light deflection of the guide star, and relativistic perigee precession of the satellite \cite{Francis1992}.)
The geodetic effect is a precession of the rotation axis of a spinning object associated with the contraction or ``warping'' of space near a massive body.
Frame-dragging is an additional precession that occurs if the massive body is rotating, pulling spacetime (and hence the gyroscope) around with it.
Popular depictions of these phenomena often model spacetime as a distorted two-dimensional sheet, but such a picture has significant limitations.
Figure~1 is an attempt at a three-dimensional view in which gravitational redshift (warping of time) is represented by a change in color of the grid lines near the Earth, the geodetic effect (warping of space) by a ``squeezing together'' of grid lines near the Earth, and the frame-dragging effect by a circular ``twisting'' of these lines in the equatorial plane.
\begin{figure}[t!]
\begin{center}
\includegraphics[width=\columnwidth]{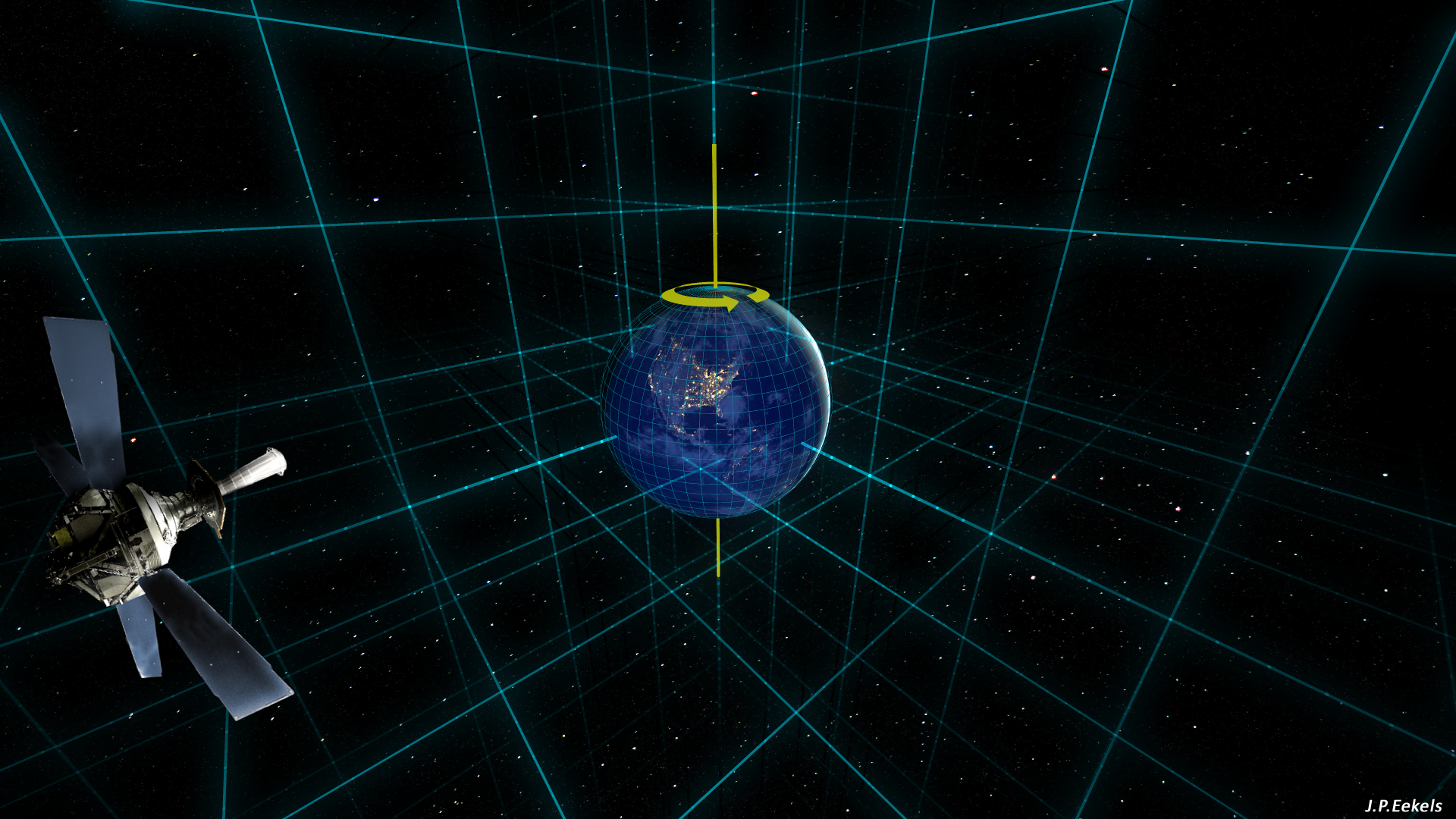}\\
\end{center}
\begin{center}
\includegraphics[width=\columnwidth]{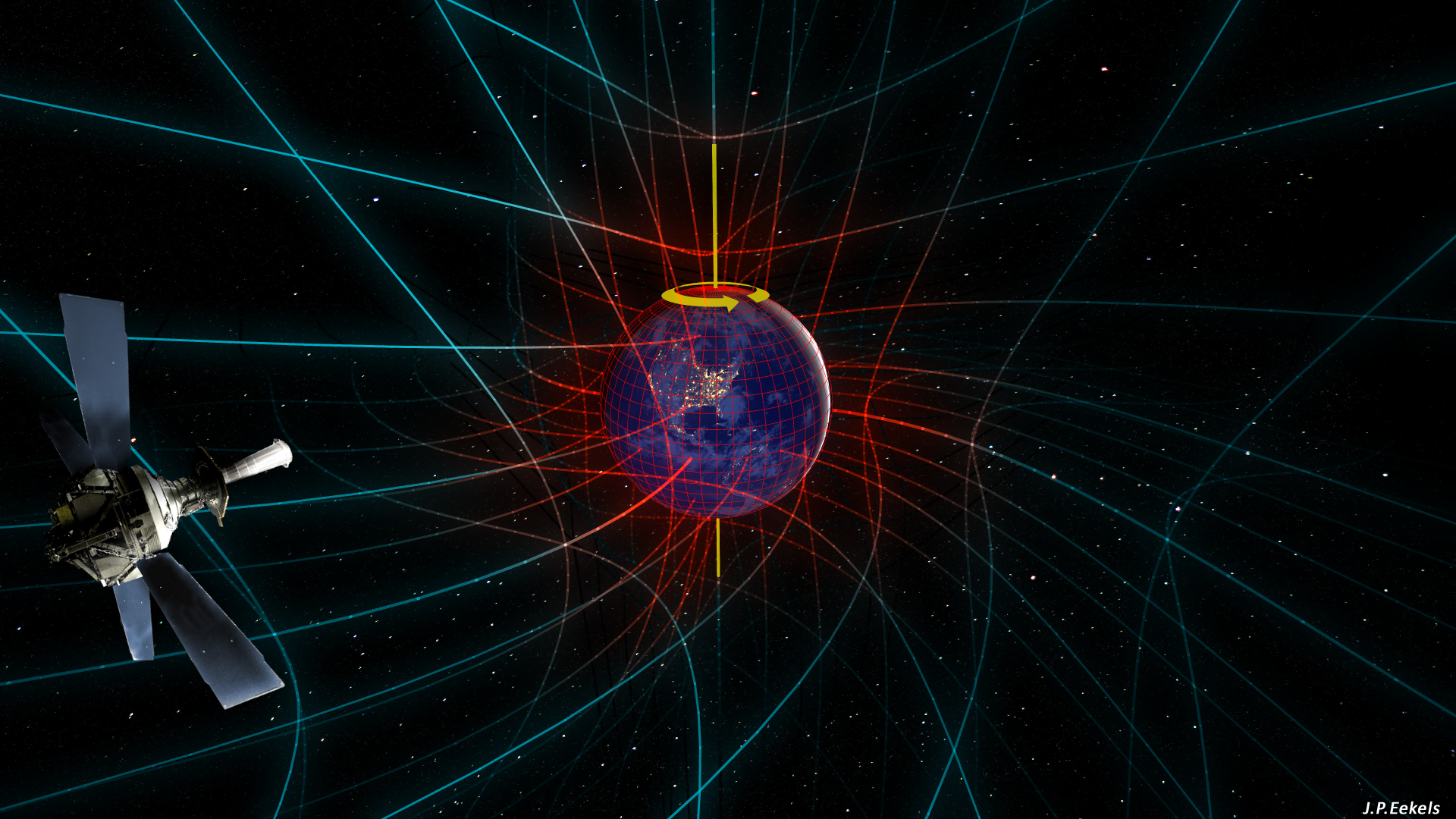}
\caption{Attempt at a schematic illustration of the differences between spacetime in Newtonian (top) and Einstein gravity (bottom) in the field of a massive, rotating body.
The warping of time (gravitational redshift) and space (geodetic effect) are represented respectively by the change in color and radial contraction of gridlines near the Earth, while the twisting of spacetime (frame-dragging effect) is represented by the spiral shape of these gridlines in the Earth's equatorial plane.
%GP-B is shown with its telescope pointed 17$^{\circ}$ above this plane.
Effects not shown to scale! [Figures courtesy J~P~Eekels, \copyright\ 2015]}
\label{fig:spacetime}
\end{center}
\end{figure}

As we know, GP-B has confirmed both predictions of GR: the geodetic effect to a precision of 0.3\% and frame-dragging to 20\% \cite{Everitt2011}.
These results constrain a class of alternatives to GR known as metric theories (loosely speaking, theories that differ from Einstein's but still respect the EP).
These are characterized by ten parametrized post-Newtonian or PPN parameters, including $\gamma$ (representing space curvature) %, $\beta$ (representing the nonlinearity of the time component of the metric) 
and a ``preferred-frame'' parameter $\alpha_1$ (allowing for possible dependence on motion relative to the mean rest frame of the Universe).
Geodetic precession is proportional to $1+2\gamma$ while frame-dragging depends on the combination $1+\gamma+\alpha_1/4$.
In GR, $\gamma=1$ and $\alpha_1=0$.
%It was originally hoped that a GP-B measurement of the geodetic effect could constrain the value of $\gamma$ to 1 part in $10^5$ \cite{Everitt1994}, equivalent to the present-day limit from Cassini.
%The need to model unexpected systematic effects (described elsewhere in this volume) has meant that subsequently weakened the GP-B bound on $\gamma$ to 4 parts in $10^3$.
Thus the GP-B measurement of geodetic precession constrains any deviation of $\gamma$ from its GR value to less than 0.4\%.

Frame-dragging is a much smaller effect, and its measurement does not usefully constrain the value of $\gamma$.
(The parameter $\alpha_1$ is already known from other tests to be smaller than $\sim10^{-4}$ \cite{Shao2012}, so the GP-B result translates into a 40\% upper limit on any deviation of $\gamma$ from its GR value).
Similar comments apply to an earlier detection of a different form of frame-dragging using the LAGEOS satellites \cite{Ciufolini1998,Ciufolini2004}.
The precession in this case involved orbital, rather than spin angular momentum, a phenomenon also known as the de~Sitter effect \cite{Overduin2010}.
The analysis assumed the GR value for geodetic precession, and its precision was limited by competing Newtonian effects due to the nonsphericity of the Earth.
The use of improved earth gravity models from the CHAMP and GRACE missions have  reduced the uncertainty in this result to approximately 10\% \cite{Ciufolini2006,Will2011}, and it is anticipated to come down further with additional data from a third satellite, LARES \cite{Ciufolini2013}.
Nevertheless, viewed as a tool for discriminating between metric theories of gravity, spin-based tests and frame-dragging in particular have sometimes been dismissed as being of little practical interest.

The importance of the geodetic and frame-dragging effects, however, does not lie in their implications for the PPN parameters.
It lies in the fact that both these phenomena are qualitatively different from all preceding tests of GR, in that they depend on the {\it spin} of the test body and/or source of the field.
The nature of this relationship between spin and spacetime within metric theories is explored in detail elsewhere in this volume by R~J~Adler.
Here we briefly discuss some aspects of what is known as the gravito-electromagnetic (GEM) analogy.
When the gravitational field is weak and velocities are small compared to $c$, then it becomes feasible to perform a $3+1$~split and decompose the metric tensor of four-dimensional spacetime into a ``gravitoelectric'' scalar potential and a  ``gravitomagnetic'' vector potential, so named because the fields obtained from the gradient and curl of these potentials obey equations that are nearly identical to Maxwell's equations and the Lorentz force law of ordinary electromagnetism \cite{Jantzen1992,Moore2012}.
The analogy is not unexpected: gravitomagnetism arises from spinning mass, just as electromagnetism arises from spinning charge.
Pushing the analogy perhaps too far, one might even describe frame-dragging of the Lense-Thirring type as gravito-paramagnetic, and that of the de~Sitter type as gravito-diamagnetic.

The terms ``frame-dragging'' and ``gravitomagnetism'' are often used interchangeably, but such an identification must be used with care, and can be as frame-dependent in gravity as it is in Maxwell's theory.
It has, for instance, been argued that gravitational tests employing lunar laser ranging (LLR) were already sensitive to frame-dragging in the Earth-Moon system before LAGEOS and GP-B \cite{Nordtvedt1988,Murphy2007}.
Others have stressed the coordinate dependence of such an assertion \cite{Kopeikin2007} and arrived at different numerical conclusions \cite{Soffel2008,Iorio2009}.
The ensuing debate has been useful in highlighting the difference between gravitoelectromagnetic effects in general, which are frame-dependent, and those that might be termed {\it intrinsically} gravitomagnetic, with an irreducible dependence on the rotation of the source of the field in appropriately defined invariants such as the contraction of the Riemann tensor with its dual \cite{Ciufolini1995}. % (p~355)
By this standard, frame-dragging (including both the Lense-Thirring and de~Sitter effects) is intrinsically gravitomagnetic, while geodetic precession and the perturbation of the lunar orbit probed by LLR are not \cite{Ciufolini2010,Pfister2012}.
It is one thing to observe an electric current and to use one's knowledge of relativistic electrodynamics to deduce that there must also exist something like a magnetic field.
It is another to detect that magnetic field {\it directly}, which is what LAGEOS and GP-B have done, in different and complementary ways.

Going beyond pure metric theories, it is possible that one would have a theory whose solutions satisfied the first five (non-spin-based) tests of GR, but not the sixth and seventh.
It is even possible (however contrived this might appear from the PPN perspective) that such a theory might agree with Einstein's prediction for geodetic precession, but not frame-dragging---or vice versa.
(One could still test such theories by associating separate phenomenological parameters with each effect, as noted by C~M~Will \cite{Will1994}.)
%To quote C~M~Will again in his commentary on the GP-B results, ``While the results support Einstein, this didn't have to be the case'' \cite{Will2011}.
A historical parallel is useful here \cite{Francis1992}.
There were similar problems with unification at the end of the 19th century, when light was known to behave one way and the rest of physics another.
The eminent physicist J~J~Thomson responded to the crisis by constructing a PPN-like framework parametrizing eight different alternatives to Maxwell's electromagnetic theory, in the hopes that experiment would gradually favor one over the others.
That framework, and all the alternative theories it encompassed, are now completely forgotten.
Instead, as we know, apparently unrelated experiments pushed the limits of existing technology in new directions, stimulating people to ask entirely new questions, and the rest is history.
Phenomenological classification schemes are of great value in comparing and testing alternatives to existing theories, but we should not allow them to unduly restrict the range of our thinking or the direction of our experiments---especially on the most fundamental issues.
% bring up this point again in connection with the SME below?
Examples like parity violation and dark energy remind us of nature's capacity to confound even near-universal consensus about what we will find.

\section{The Standard Model}

No consensus has yet emerged about the best way to combine gravity with the other fundamental interactions of nature.
Nevertheless, there have long been reasons to suspect that any such unification will involve spin in some way.

Spin plays an explicit dynamical role in non-Riemannian extensions of GR that endow spacetime with torsion, a possibility that Einstein chose to ignore for reasons of logical economy, but which arguably adds to the formal beauty of his theory \cite{Hehl1976}.
Possible implications of some torsion-based theories for GP-B were discussed by L~Halpern beginning in the 1980s \cite{Leo1984} (see for review \cite{OP2008}), and more recently by Y~Mao et al. \cite{Mao2007}.
However, the current consensus is that torsion in the most natural theories (based on Poincar\'e gauge symmetry) couples only to microscopic, not macroscopic spin \cite{Flanagan2007a,Hehl2013}, and would only be detectable in principle using polarized test materials \cite{Ni2010}.
Alternatively, the existing results from GP-B can be interpreted as constraints on a generalized and perhaps more fundamental version of the EP in which gravitation is rotationally, as well as translationally equivalent to acceleration \cite{Ni2011}.

Torsion is associated mathematically with antisymmetric terms in the affine connection (Christoffel symbol) of GR, so it is not surprising that theories which allow the metric itself to be nonsymmetric can be tested in a similar way.
GP-B constraints on such theories, which agree with the GR prediction for frame-dragging but not for geodetic precession, have been discussed by J~W~Moffat \cite{Moffat1990}.

A different approach is to supplement the metric tensor of GR with one or more scalar fields of the kind that appear in particle physics in the context of spontaneous symmetry-breaking, and also in many approaches to unification, including string theories.
In the 1990s, work by T~Damour and K~Nordtvedt \cite{Damour1993} showing that one such class of theories evolves generically toward GR led to renewed interest in experimental bounds on the PPN parameters.
More recently, GP-B constraints on scalar-tensor theories have been discussed by K~K~Nandi et al. \cite{Nandi2001}.

A widely followed approach to unification involves extending the $(3+1)$-dimensional spacetime of Einstein's theory to $N+1$ dimensions, as in Kaluza-Klein theories and their descendants, string and brane theories.
A major attraction of these theories
%(like those based on torsion or nonsymmetric metric terms)
is their potential for preserving what is known as ``Einstein's dream'': the idea that all matter and energy in the four-dimensional world is a manifestation of pure geometry \cite{Wesson1996}.
The new degrees of freedom in the metric tensor can be reformulated as scalar fields, so there is some overlap between this approach and the one just discussed.
But spin enters into the discussion in a particularly compelling way here, because it is generally recognized that greatest obstacle to the geometrization of standard-model fields via higher dimensions is posed by particles with nonzero spin \cite{OW1997}.
Constraints on Kaluza-Klein gravity from geodetic precession have been worked out %for the case of one additional dimension
by D~Kalligas et al. \cite{KWE1995} and H~Liu et al. \cite{LW1996,LO2000} and updated in light of the GP-B results by J~M~Overduin, R~D~Everett and P~S~Wesson \cite{OEW2013}.

A more phenomenological approach to the problem of unification has led to the creation of the Standard-Model Extension (SME), a PPN-like framework that parametrizes possible violations of Lorentz symmetry in all of known physics and encompasses both GR and the SM as special cases.
The full SME contains hundreds of parameters, but its ``pure-gravity sector'' involves just nine, corresponding to the vacuum expectation values of a new tensor field (assumed constant in the asymptotically flat or Minkowski limit) whose couplings to the traceless part of the Ricci tensor induce spontaneous violations of local Lorentz symmetry \cite{Kostelecky2004}.
The geodetic and frame-dragging effects are sensitive to two different combinations of seven of these SME coefficients \cite{BK2006}.
% The constraint from frame-dragging turns out to be superseded by a linear combination of constraints from existing experiments, but that from geodetic precession is unique. 
GP-B constraints on the SME have been calculated by Q~B~Bailey, R~D~Everett and J~M~Overduin and include the first upper bound on the magnitude of the SME time-time coefficient \cite{BEO2013}.
(This has since been strengthened by an elegant argument employing boost transformations and binary pulsars \cite{Shao2014}.)
The SME allows for precession rates larger than that predicted by GR, in contrast to theories based on scalar fields or extra dimensions, for which GR acts as a limiting case.

In the three years since they appeared, the results from GP-B have led to new limits on a variety of other fundamental physics theories, most of them motivated in one way or another by the challenge of unification.
Examples include Cherns-Simons gravity \cite{Alexander2007,AliHaimoud2011},
conformal gravity \cite{LeviSaid2013},
pseudo-complex GR \cite{Schonenbach2013},
the Tensor-Vector-Scalar theory (TeVeS) that developed as a relativistic generalization of Modified Newtonian Dynamics or MOND \cite{Exirifard2013},
noncommutative spectral geometry \cite{Lambiase2013},
Horava-Lifshitz gravity \cite{Radicella2014}, and
Extended Gravity, in which the action is allowed to depend on a general function of the Ricci scalar \cite{Capozziello2014}.

To sum up, it is fair to say that no one alternative to GR has yet gained wide acceptance as a promising approach to unification of gravity with the SM.
There is some consensus that metric theories will not suffice, but it is not clear whether any of the non-metric alternatives discussed above are ``crazy enough to have a chance of being correct'' (to quote N~Bohr).
The GP-B results have not yet proved decisive in this regard, but may well do so in the years to come.
%If history is any guide, the best way to stimulate further progress on the theoretical side may be to dig down further on the experimental side, to the foundation that underlies nearly all our ideas about gravity: the EP.

\section{Astrophysics}

By directly verifying the reality of the geodetic and frame-dragging effects for spinning bodies, LAGEOS and GP-B have strengthened our confidence that we understand the mechanisms underlying some of the remotest and most energetic phenomena in the Universe \cite{Thorne1988}.
Prime among these are the vast jets of gas and magnetic field ejected from quasars and active galaxies like the radio source NGC~6251.
These jets are fed by material from the accretion disks surrounding supermassive black holes at their cores.
The Mpc length scale of the jets implies that their direction is held
constant over time scales as long as tens of millions of years.
This is thought to be accomplished by the gyroscopic spin of the black hole, whose direction is communicated to the jet via the black hole's gravitomagnetic field.
Frame-dragging causes the disk to precess around the black hole, and this precession combines with the disk's viscosity to drive the inner region of the disk into the black hole's equatorial plane, so that the jets gradually become aligned with the spin axis of the black hole.
This phenomenon, known as the Bardeen-Petterson effect \cite{Bardeen1975,Kumar1985}, has been verified by detailed three-dimensional numerical simulations \cite{Nelson2000,Nealon2015}.

Gravitoelectromagnetism is also thought to lie behind the generation of the astounding energy contained in these jets in the first place.
Interaction between the gravitomagnetic potential of the black hole and infalling plasma from the magnetized accretion disk effectively transforms the horizon into a gigantic ``gravitational battery'' that converts the hole's immense rotational energy into a stream of outgoing ultra-relativistic charged particles \cite{Thorne1988}.
This mechanism, known as the Blandford-Znajek process \cite{Blandford1977,MacDonald1982}, has also been validated in numerical simulations \cite{Komissarov2001,Penna2013}.

Frame-dragging may help us understand the interiors of neutron stars \cite{Blandford1994,Levin2004} and can likely explain certain kinds of quasi-periodic oscillations in accretion disks around black holes \cite{Silbergleit2001} that have now been seen at both high and low frequencies \cite{Ingram2009,Stefanov2014}.
Geodetic precession has been confirmed in binary pulsar systems such as B1534+12 \cite{Stairs2004} and J1141-6545 \cite{Hotan2005}, and implicated in the ``disappearance'' of the pulsar J1906+0746 as its spin axis rotates out of the line of sight \cite{vanLeeuwen2015} due to the curvature of space generated by its more massive companion.
The expected advent of gravitational-wave detection will open another window on gravitomagnetism through its effects on the orbits of neutron-star and black-hole binaries in the final stages of inspiral and coalescence \cite{Flanagan2007b}.
Perhaps most remarkably, frame-dragging is now becoming part of the  toolkit for cosmological simulations of large-scale structure \cite{Bruni2014}.

\section{Cosmology}

It is on cosmological scales that frame-dragging may take on its
deepest significance, as part of the explanation for the origin of inertia.
The Lense-Thirring effect whose existence has been confirmed by LAGEOS and GP-B  involves the dragging of inertial frames outside a massive, rotating sphere.
The cosmological counterpart to this effect, also studied by Thirring in 1917, involves the dragging of local inertial frames inside a massive, rotating shell.
(Historical work has now established that Einstein provided the key insights in both derivations \cite{Pfister2007}.)
Specifically, Thirring showed that an observer near the center of a spherical shell of radius $R$ and mass $M$ rotating with an angular velocity $\vec{\omega}$ experiences an acceleration given by $-2d_1(\vec{\omega}\times\vec{v})-d_2[\vec{\omega}\times\vec{\omega}\times\vec{r}+2(\vec{\omega}\cdot\vec{r})\vec{\omega}]$ where $\vec{r}$ and $\vec{v}$ are the observer's position and velocity, and where (to first order in $r/R$) $d_1=4GM/3c^2R$ and $d_2=4GM/15c^2R$.
When $d_1=d_2=1$ this is just the sum of the Coriolis and centrifugal accelerations experienced in a rotating reference frame in standard Newtonian mechanics (plus an ``anomalous'' extra term proportional to $\vec{\omega}$; see below).  
Rough estimates do in fact suggest that $d_1\sim d_2\sim 1$ \cite{Sciama1959}.
That is, if the Universe were really to rotate around us in this manner, we would co-rotate with it.
Known as ``perfect dragging,'' this result suggests that rotational inertia may be a manifestation of frame-dragging within GR, rather than a measure of motion relative to absolute space, as in Newtonian theory.
 
Thirring's model was overly simplistic but the phenomenon of perfect dragging has persisted through a series of increasingly sophisticated treatments \cite{Pfister1995}.
D~R~Brill and J~M~Cohen obtained the same result using a Schwarzschild background, but with a Coriolis coefficient $d_1=4\alpha(2-\alpha)/[(1+\alpha)(3-\alpha)]$ where $\alpha=GM/2c^2R$ \cite{Brill1966}.
It is suggestive (if undoubtedly coincidental) that $d_1=1.0$ for a closed Universe with $M=\rho V$, $V=2\pi^2R^3$, $R=ct_0$ and $\rho=\rho_{\mbox {\tiny crit}}=3H_0^2/8\pi G$, $H_0t_0=1.0$ as observed \cite{Wilcomb2014}.
L~Lindblom and D~R~Brill generalized further to include time dependence, and emphasized that inertial dragging in this picture occurs {\it instantaneously} without violating causality  \cite{Lindblom1974}.
(This important but subtle point has been discussed by many people \cite{MTW73,Ciufolini1995,LyndenBell1996,Katz1998}; the key is that the determination of spacetime geometry takes place on a spacelike hypersurface.)
H~Pfister and K~Braun took the calculation to higher orders in rotational velocity and concluded that the centrifugal term also fulfills Newtonian expectations (thereby addressing the above-mentioned shortcoming of Thirring's original model) \cite{Pfister1986}.
Further refinements have involved more realistic FRW-type cosmological models \cite{Klein1993,Klein1994}, the gyromagnetic ratio of a charged rotating mass shell \cite{Pfister2002} and  cosmological perturbations \cite{Schmid2006,Schmid2009,Schmid2014}.
Evidence has also been found for perfect dragging in the {\it translational} sense \cite{Gron1989,Pfister2005}.

Some have raised objections to the apparent elimination of absolute spacetime in this way.
B~Mashhoon has argued forcefully against dragging-type explanations for Newtonian centrifugal force in particular \cite{Mashhoon1993} and translational inertia in general \cite{Mashhoon1994}.
W~Rindler has shown that paradoxical issues arise if one interprets gravitomagnetic effects too literally in terms of the ``space dragging'' \cite{Rindler1997}.
There is a mismatch between the gravitational and electromagnetic gyromagnetic ratios for charged fermions at the quantum-mechanical level \cite{Adler2012}.
It is possible to overstate the apparent conflict between the absolute and relational viewpoints; both are likely to appear as complementary aspects of any eventual unified theory when gravitational and other degrees of freedom are taken into account as part of the ``matter and energy content'' of the Universe \cite{Overduin2001,Barbour2012}.
Such a theory may need to be formulated on the assumption of a finite (spatially closed) Universe if it is to account for inertia in a self-consistent way \cite{Barbour2010}.
Despite a widespread view to the contrary, such a possibility is  fully compatible with observational cosmology and likely to remain so for the forseeable future \cite{Adler2005}.

%These cosmological considerations, like the astrophysical ones in the preceding section, do not, of course, bear {\it directly} on the GP-B results.
%But they are of great relevance {\it indirectly}, in the same sense that laboratory experiments on Earth are relevant to our understanding of astrophysics and cosmology.
%We trust our theories of stellar fusion, in part, because of our experience with terrestrial $\beta$-decay.
%In the same way, the first direct observation of gyroscopic frame-dragging gives us hope that we will one day understand the even more fundamental question of the origin of inertia.

\section{Summary}

There are more reasons now than ever before to conduct gravitational tests of the greatest possible range and sensitivity.
This necessarily incomplete review has focused on existing spin-based tests and their constraints on non-metric theories, as motivated particularly by the problem of unification with the standard model of particle physics.
In the three years since they have been published, measurements of the geodetic and frame-dragging effects by Gravity Probe~B have led to new constraints on at least ten such theories or phenomenological frameworks.
The results have not yet proved decisive, but that may reflect less on the experiment and more on the current state of theory in the field.
The GP-B and LAGEOS results also increase our confidence that we correctly understand the theoretical mechanisms governing astrophysical jets, neutron-star and black-hole binary systems, and perhaps even the origin of inertia itself.

Any effort as ambitious as GP-B is inevitably followed by questions.
In some respects the situation is reminiscent of that in the years following the Apollo program.
It is possible to argue that much of the same knowledge could have been gained in other ways; we already knew what the answers were likely to be.
It is also possible to argue that to push existing technology to the limit on an issue so fundamental is itself a worthy goal.
Robotic sample return missions could have explored the Moon for us.
But the fact remains: we have walked on another world.
The gravito-electromagnetic aspects of Einstein's theory of General Relativity can now be explored in ways unforeseen when GP-B was first conceived.
But the fact remains: we have spun tops, and through them, walked metaphorically on the warped and twisted fabric of spacetime itself.

\ack
It is a pleasure to thank Ron Adler, Julian Barbour, Francis Everitt, Bahram Mashhoon, Alex Silbergleit and Paul Wesson for friendly discussions.
Support is acknowledged from the Faculty Development and Research Committee of Towson University.


\section*{References}

\begin{thebibliography}{99}
\bibitem{STEP} Overduin J M, Everitt C W F, Worden P and Mester J 2012 {\it Class. Quant. Grav.} {\bf 29} 184012
\bibitem{Will} Will C M 2014 {\it Living Rev. Relativity} {\bf 17} 4
\bibitem{Gwaves} Aasi J et al. 2013 {\it Phys. Rev. D} {\bf 87} 022002
\bibitem{Everitt2011} Everitt C W F et al. 2011 {\it Phys. Rev. Lett.} {\bf 106} 221101
\bibitem{Shao2012} Shao L and Wex N 2012 {\it Class. Quant. Grav.} {\bf 29} 215018
%\bibitem{Francis1992} Everitt C W F 1991 in Sato~H and Nakamura~T (eds), proc. Sixth Marcel Grossmann meeting on recent developments in theoretical and experimental General Relativity, gravitation and relativistic field theories (Singapore: World Scientific, 1992)~1632
%\bibitem{Everitt1994} Everitt C W F and Buchman S 1994, in Tr\^an Thanh V\^an~J, Fontaine~G and Hinds~E (eds), {\it Particle astrophysics, atomic physics and gravitation} (Singapore: \'Editions Fronti\`eres, 1994)~467
% LAGEOS/LARES
\bibitem{Ciufolini1998} Ciufolini I, Pavlis E C, Chieppa F, Fernandes-Vierira E and P\'erez-Mercader J 1998 {\it Science} {\bf 279} 2100
\bibitem{Ciufolini2004} Ciufolini I and Pavlis E C 2004 {\it Nature} {\bf 431} 958
\bibitem{Overduin2010} Overduin J M 2010 in Petkov V (ed) {\it Space, Time and Spacetime} (Berlin: Springer) 25
\bibitem{Ciufolini2006} Ciufolini I, Pavlis E C and Peron R 2006 {\it New Astron.} {\bf 11} 527
\bibitem{Will2011} Will C M 2011 {\it Physics} {\bf 4} 43
\bibitem{Ciufolini2013} Ciufolini I, Moreno Monge B, Paolozzi A, Koenig R, Sindoni G, Michalak G and Pavlis E C 2013 {\it Class. Quant. Grav.} {\bf 14} 2701
% Gravitomagnetism
\bibitem{Jantzen1992} Jantzen R T, Carini P and Bini D 1992 {\it Ann. Phys.} (NY) {\bf 215} 1
\bibitem{Moore2012} Moore T A 2012 {\it A General Relativity Workbook} (University Science Books) 405
% Coordinate confusion
\bibitem{Nordtvedt1988} Nordtvedt K 1988 {\it Int. J. Theor. Phys.} {\bf 27} 1395
\bibitem{Murphy2007} Murphy T W, Nordtvedt K and Turyshev S G 2007 {\it Phys. Rev. Lett.} {\bf 98} 071102 %; arXiv:gr-qc/0702028
\bibitem{Kopeikin2007} Kopeikin S M 2007 {\it Phys. Rev. Lett.} {\bf 98} 229001 %; arXiv:gr-qc/0702120
\bibitem{Soffel2008} Soffel M, Klioner S, M\"uller J and Biskupek L 2008 {\it Phys. Rev. D} {\bf 78} 024033
\bibitem{Iorio2009} Iorio L 2009 {\it Int. J. Mod. Phys. D} {\bf 18} 1319 %; arXiv:0809.4014
% Resolution using curvature invariants
\bibitem{Ciufolini1995} Ciufolini I and Wheeler J A 1995 {\it Gravitation and Inertia} (Princeton Univ. Press)
\bibitem{Ciufolini2010} Ciufolini I 2010 {\it New Astron.} {\bf 15} 332 %; arXiv:0809.3219
\bibitem{Pfister2012} Pfister H 2012 {\it Gen. Rel. Grav.} {\bf 44} 3217
% Metric or non-metric
\bibitem{Will1994} Will C M 1994 {\it Phys. Rev. D} {\bf 67} 062003
\bibitem{Francis1992} Everitt C W F 1992 in Sato~H and Nakamura~T (eds) {\it Sixth Marcel Grossmann Meeting on Recent Developments in Theoretical and Experimental General Relativity, Gravitation and Relativistic Field Theories} (Singapore: World Scientific)~1632
% Torsion
\bibitem{Hehl1976} Hehl F W, von der Heyde P, Kerlick G D and Nester J M 1976 {\it Rev. Mod. Phys.} {\bf 48} 393
\bibitem{Leo1984} Halpern L 1984 {\it Int. J. Theor. Phys.} {\bf 23} 843
\bibitem{OP2008} Overduin J and Plendl H 2008, in Kleinert~H, Jantzen~R~T and Ruffini~R (eds) {\it Eleventh Marcel Grossmann Meeting on General Relativity} (Singapore: World Scientific)~870
\bibitem{Mao2007} Mao Y, Tegmark M, Guth A and Cabi S 2007 {\it Phys. Rev. D} {\bf 76} 104029
\bibitem{Flanagan2007a} Flanagan \'E \'E and Rosenthal E 2007 {\it Phys. Rev. D} {\bf 75} 124016
\bibitem{Hehl2013} Hehl F W, Obukhov Y N and Puetzfeld D 2013 {\it Phys. Lett. A} {\bf 377} 1775
\bibitem{Ni2010} Ni W-T 2010 {\it Rep. Prog. Phys.} {\bf 73} 056901
\bibitem{Ni2011} Ni W-T 2011 {\it Phys. Rev. Lett.} {\bf 107} 051103
% Nonsymmetric gravity
\bibitem{Moffat1990} Moffat J W and Brownstein J R 1990 {\it Phys. Rev. D} {\bf 41} 3111; Moffat J W 2004 arXiv:gr-qc/0405091
% Scalar fields
\bibitem{Damour1993} Damour T and Nordtvedt K 1993 {\it Phys. Rev. Lett.} {\bf 70} 2217
\bibitem{Nandi2001} Nandi K K, Alsing P M, Evans J C and Nayak T B 2001 {\it Phys. Rev. D} {\bf 63} 084027
% Kaluza-Klein
\bibitem{Wesson1996} Wesson P S et al. 1996 {\it Int. J. Mod. Phys. A} {\bf 11} 3247
\bibitem{OW1997} Overduin J M and Wesson P S 1997 {\it Phys. Rep.} {\bf 283} 
\bibitem{KWE1995} Kalligas D, Wesson P S and Everitt C W F 1995 {\it Astrophys. J.} {\bf 439} 548
\bibitem{LW1996} Liu H and Wesson P S 1996 {\it Class. Quant. Grav.} {\bf 13} 2311
\bibitem{LO2000} Liu H and Overduin J M 2000 {\it Astrophys. J.} {\bf 538} 386
\bibitem{OEW2013} Overduin J M, Everett R D and Wesson P S 2013 {\it Gen. Rel. Grav.} {\bf 45} 1723
% Standard-Model Extension
\bibitem{Kostelecky2004} Kosteleck\'y V A 2004 {\it Phys. Rev. D} {\bf 69} 105009
\bibitem{BK2006} Bailey Q G and Kosteleck\'y V A 2006 {\it Phys. Rev. D} {\bf 74} 045001
\bibitem{BEO2013} Bailey Q G, Everett R D and Overduin J M 2013 {\it Phys. Rev. D} {\bf 88} 102001
\bibitem{Shao2014} Shao L 2014 {\it Phys. Rev. D} {\bf 90} 122009
% Cherns-Simons gravity
\bibitem{Alexander2007} Alexander S and Yunes N 2007 {\it Phys. Rev. Lett.} {\bf 99} 241101
\bibitem{AliHaimoud2011} Ali-Ha\"imoud Y and Chen Y 2011 {\it Phys. Rev. D} {\bf 84} 124033
% Conformal gravity
\bibitem{LeviSaid2013} Said J L, Sultana J and Adami K Z 2013 {\it Phys. Rev. D} {\bf 88} 087504
% pseudo-complex GR
\bibitem{Schonenbach2013} Sch\"onenbach T, Caspar G, Hess P O, Boller T, M\"uller A, Sch\"afer M and Greiner W 2013 {\it Mon. Not. R. Astron. Soc.} {\bf 430} 2999
% TeVeS
\bibitem{Exirifard2013} Exirifard Q 2013 {\it J. Cosmology Astropart. Phys.} {\bf 04} 034 %; arXiv: 1111.5210
% non-commutative spectral action
\bibitem{Lambiase2013} Lambiase G, Sakellariadou M and Stabile An 2013 {\it J. Cosmology Astropart. Phys.} {\bf 12} 020 %; arXiv: 1302.2336
% Horava-Lifshitz gravity
\bibitem{Radicella2014} Radicella N, Lambiase G, Parisi L and Vilasi G 2014 {\it J. Cosmology Astropart. Phys.} {\bf 12} 014 %; arXiv: 1408.1247
% Extended f(R) gravity
\bibitem{Capozziello2014} Capozziello S, Lambiase G, Sakellariadou M, Stabile An and Stabile Ar 2014, arXiv:1410.8316
% Astrophysics
\bibitem{Thorne1988} Thorne K 1988 in Fairbank J~D, Deaver~B~S, Everitt, C~W~F and Michelson P~F (eds) {\it Near Zero: New Frontiers of Physics} (New York: W.H. Freeman)~572
% Bardeen-Petterson
\bibitem{Bardeen1975} Bardeen J M and Petterson J A 1975 {\it Astron. Astrophys. Lett.} {\bf 195} L65
\bibitem{Kumar1985} Kumar S and Pringle J E 1985 {\it Mon. Not. R. Astron. Soc.} {\bf 213} 435
\bibitem{Nelson2000} Nelson R P and Papaloizou J C B 2000 {\it Mon. Not. R. Astron. Soc.} {\bf 315} 570
\bibitem{Nealon2015} Nealon R, Price D J and Nixon C J 2015 arXiv:1501.0168
% Blandford-Znajek
\bibitem{Blandford1977} Blandford R D and Znajek R L 1977 {\it Mon. Not. R. Astron. Soc.} {\bf 179} 433
\bibitem{MacDonald1982} MacDonald D and Thorne K S 1982 {\it Mon. Not. R. Astron. Soc.} {\bf 198} 345
\bibitem{Komissarov2001} Komissarov S S 2001 {\it Mon. Not. R. Astron. Soc.} {\bf 326} L41
\bibitem{Penna2013} Penna R F, Narayan R and Sadowski 2013 {\it Mon. Not. R. Astron. Soc.} {\bf 436} 3741
% Frame-dragging and neutron stars
\bibitem{Blandford1994} Blandford R D 1995 {\it J. Astrophys. Astr.} {\bf 16} 191
\bibitem{Levin2004} Levin Y and D'Angelo C 2004 {\it Astrophys. J.} {\bf 613} 1157
% Frame-dragging and QPOs in X-ray binaries
\bibitem{Silbergleit2001} Silbergleit A S, Wagoner R V and Ortega-Rodr\'iguez M 2001 {\it Astrophys. J.} {\bf 548} 335
\bibitem{Ingram2009} Ingram A, Done C and Fragile P C 2009 {\it Mon. Not. R. Astron. Soc.} {\bf 397} L101
\bibitem{Stefanov2014} Stefanov I Z 2014 {\it Mon. Not. R. Astron. Soc.} {\bf 444} 2178
% Geodetic precession in pulsars
\bibitem{Stairs2004} Stairs I H, Thorsett S E and Arzoumanian Z 2004 {\it Phys. Rev. Lett.} {\bf 93} 141101
\bibitem{Hotan2005} Hotan A W, Bailes M and Ord S M 2005 {\it Astrophys. J.} {\bf 624} 906
\bibitem{vanLeeuwen2015} van Leeuwen J et al. 2015 {\it Astrophys. J.} {\bf 798} 118
% Frame-dragging signatures in gravitational waves
\bibitem{Flanagan2007b} Flanagan \'E \'E and Racine \'E 2007 {\it Phys. Rev. D} {\bf 75} 044001
% Frame-dragging in cosmological simulations!
\bibitem{Bruni2014} Bruni M, Thomas D B and Wands D 2014 {\it Phys. Rev. D} {\bf 89} 044010
% Cosmology: Thirring
\bibitem{Pfister2007} Pfister H 2007 {\it Class. Quant. Grav.} {\bf 39} 1735
\bibitem{Sciama1959} Sciama D 1959 {\it The Unity of the Universe} (London: Faber and Faber)
\bibitem{Pfister1995} Pfister H in Barbour~J and Pfister~H (eds) {\it Mach's Principle: From Newton's Bucket to Quantum Gravity} (Berlin: Birkh\"auser)~315
\bibitem{Brill1966} Brill~D~R and Cohen~J~M 1966 {\it Phys. Rev.} {\bf 143} 1011
\bibitem{Wilcomb2014} Wilcomb K and Overduin~J~M 2014 {\it Bull.  Am. Phys. Soc.} {\bf 59} F1.00047
\bibitem{Lindblom1974} Lindblom~L and Brill~D~R 1974 {\it Phys. Rev.} {\bf D10} 3151
% Instantaneous!
%\bibitem{Ciufolini1995} Ciufolini I and Wheeler J A 1995 {\it Gravitation and Inertia} (Princeton Univ. Press) 274
\bibitem{MTW73} Misner C W, Thorne K S and Wheeler J A 1973 {\it Gravitation} (New York: W H Freeman) % p~543
\bibitem{LyndenBell1996} Lynden-Bell D 1996 in Lahav O, Terlevich E and Terlevich~R~J (eds) {\it Gravitational Dynamics} (Cambridge Univ. Press) 235
\bibitem{Katz1998} Katz J, Lynden-Bell~D and Bi\v{c}\'{a}k~J 1998 {\it Class. Quant. Grav.} {\bf 15} 3177
% More realistic models
\bibitem{Pfister1986} Pfister H and Braun K 1985 {\it Class. Quant. Grav.} {\bf 2} 909
\bibitem{Klein1993} Klein C 1993 {\it Class. Quant. Grav.} {\bf 10} 1619
\bibitem{Klein1994} Klein~C 1994 {\it Class. Quant. Grav.} {\bf 11} 1539
% Gyromagnetic ratio
\bibitem{Pfister2002} Pfister H and King M 2002 {\it Phys. Rev. D} {\bf 65} 084033
% Perturbations
\bibitem{Schmid2006} Schmid C 2006 {\it Phys. Rev. D} {\bf 74} 044031
\bibitem{Schmid2009} Schmid C 2009 {\it Phys. Rev. D} {\bf 79} 064007
\bibitem{Schmid2014} Schmid C 2014 arXiv:1406.4665
% Linear dragging!
\bibitem{Gron1989} Gr{\o}n {\O} and Eriksen E 1989 {\it Gen. Rel. Grav.} {\bf 21} 105
\bibitem{Pfister2005} Pfister H, Frauendiener J and Hengge S 2005 {\it Class. Quant. Grav.} {\bf 22} 4743
% Challenges!
\bibitem{Mashhoon1993} Mashhoon 1993 in Hu B L and Jacobson T A (eds) {\it Directions in General Relativity} Vol. 2 (Cambridge Univ. Press)~182 
\bibitem{Mashhoon1994} Mashhoon 1994 {\it Il Nuovo Cim. B} {\bf 109} 187
\bibitem{Rindler1997} Rindler W 1997 {\it Phys. Lett. A} {\bf 233} 25
\bibitem{Adler2012} Adler R J, Chen P and Varani E 2012 {\it Phys. Rev. D} {\bf 85} 025016
\bibitem{Overduin2001} Overduin J M and Fahr H-J 2001 {\it Naturwissenschaften} {\bf 88} 491
\bibitem{Barbour2012} Barbour J 2012 {\it Annalen der Physik} {\bf 524} A39
\bibitem{Barbour2010} Barbour J 2010 {\it Found. Phys.} {\bf 40} 1263
\bibitem{Adler2005} Adler R J and Overduin J M 2005 {\it Gen. Rel. Grav.} {\bf 37} 1491
\end{thebibliography}
\end{document}